# Global Minimum Depth In Edwards-Anderson Model

I. Karandashev,[1,2][0000-0001-8483-072X] and B. Kryzhanovsky[1][0000-0002-0901-6370]

[1] Center of Optical Neural Technologies, Scientific Research Institute for System Analysis RAS, Moscow, Nakhimovskiy prosp., 36, b.1., 117218, Russia
[2] Peoples Friendship University of Russia (RUDN University), 6 Miklukho-Maklaya St, Moscow, 117198, Russian Federation
`{karandashev, kryzhanov}@niisi.ras.ru`

**Abstract.** In the literature the most frequently cited data are quite contradictory, and there is no consensus on the global minimum value of 2D Edwards-Anderson (2D EA) Ising model. By means of computer simulations, with the help of exact polynomial Schraudolph-Kamenetsky algorithm, we examined the global minimum depth in 2D EA-type models. We found a dependence of the global minimum depth on the dimension of the problem N and obtained its asymptotic value in the limit $N\to\infty$. We believe these evaluations can be further used for examining the behavior of 2D Bayesian models often used in machine learning and image processing.

## 1 Introduction

In many fields of science, it is necessary to know the global energy minimum for different systems. Namely, in informatics we use it when solving problems of quadratic optimization [1–6], developing search algorithms for the global minimum [7–12] and solving max-cut problems [13–17]. In neuroinformatics, we have to know the global minimum when developing associative memory systems [18–21] and constructing neural networks and neural network minimization algorithms [22–24]. In physics, the knowledge of the global energy minimum is most frequently necessary when studying the behavior of spin glass systems [25–35] and even when describing four-photon mixing in nonlinear media [36, 37].

The question of calculation of the global minimum depth has been discussed over the years. However, since it has no decisive answer it remains a highly topical problem up to now. Indeed, in the literature the most frequently cited data are quite contradictory, and there is no consensus on the global minimum value (see references in [27]). To illustrate this statement, we present the values of the global minimum depth obtained by different methods:

$$E_0 = 0 \qquad \text{TAP (Thouless et al. [25])}$$
$$E_0 = 1/\sqrt{2\pi} \qquad \text{mean random field (Klein [26])}$$
$$E_0 = 0.5 \qquad \text{partition function (Tanaka and Edwards [27])} \qquad (1)$$
$$E_0 = 2/\pi \qquad \text{replica (Sherrington and Kirkpatrick [28])}$$
$$E_0 = 0.76 \sim 0.77 \qquad \text{Monte Carlo (Sherrington and Kirkpatrick [29])}$$

Such a spread of values exists because until recently there were no exact calculation algorithms for the determination of $E_0$. This was the reason why different authors used different minimization methods, and consequently the obtained estimates were sufficiently far from the true value of $E_0$. New algorithms appeared recently. They allow us to calculate $E_0$ exactly when examining spin systems on planar graphs with arbitrary boundary conditions [38, 39]. Implementing these algorithms, we were able to refine our results [40] for the Edwards–Anderson model (the EA model).

In the present paper, we present an experimental analysis of the global minimum depth in the EA model, which is a spin system on an $N = L \times L$ square lattice where only interactions with four nearest neighbors do not equal to zero. Formally, we have in mind a system whose behavior is described by a Hamiltonian

$$H = -\frac{1}{2}\sum_{i,j=1}^{N} J_{ij} s_i s_j \qquad (2)$$

defined in the configuration space of states $\mathbf{S} = (s_1, s_2, ..., s_N)$ with binary variables $s_i = \pm 1$, $i = \overline{1, N}$. Here $N$ is the number of spins, and $J_{ij}$ is a symmetric, zero-diagonal matrix ($J_{ij} = J_{ji}$ and $J_{ii} = 0$).

To describe the spectrum of the system, it is convenient to introduce the depth of the minimum that is defined by equation

$$E = \frac{1}{2N\sigma_J} \sum_{i=1}^{N} \sum_{j=1}^{N} J_{ij} s_i s_j \qquad (3)$$

As we show in what follows, the normalization coefficient in Eq. (3) is quite universal since the value of E is almost independent of the dimension of the problem $N$ as well as of the normalization of the matrix elements $J_{ij}$. In these notations, the Hamiltonian of the system has the form $H = -N\sigma_J E$, and its dependence on the dimension reduces to $H \sim N$.

As we see from Eq. (1), the results obtained by different authors are so very different that it is hardly possible to use them in the course of calculations. This was the reason why we performed a huge experiment having in mind to determine the basic spectral characteristics such as the mean value of the local minimum depth, the spectrum width, and the depth of the global minimum. Based on the obtained experimental data, we plotted the dependences of these characteristics on the dimension of the problem $N$ and determined their asymptotic values in the limit $N \to \infty$.

The structure of the paper is as follows. In Section II, we describe our experiment and analyze the obtained data. In Section III, we discuss the results and the tables showing our experimental data.

## 2 Experiment

To define the value of $E_0$, we used an algorithm described in [39]. In the course of our experiment, we examined the classical EA-model (with the normal distribution of $J_{ij}$) and the EA*-model (with the uniform distribution of $J_{ij}$). For the chosen model of the given size $N = L \times L$, we generated $M$ matrices $J_{ij}$ and determined $M$ values $E_{m0}$, $m = \overline{1, M}$. We used these data to calculate the mean value and the variance of the obtained values:

$$E_0 = M^{-1} \sum_{m=1}^{M} E_{m0}, \quad \sigma_0^2 = M^{-1} \sum_{m=1}^{M} E_{m0}^2 - E_0^2 \qquad (4)$$

The results of our experiments are collected in Table 1.

Based on the obtained data, we derived formulas that described the dependences of $E_0$ and $\sigma_0$ on $N$. We optimized these formulas by means of the least squares method. We minimized the value of the summary relative error and estimated the quality of the approximation formulas by the value of validity defined as

$$R^2 = 1 - \frac{\sum (x_{\exp} - x_{approx})^2}{\sum (x_{\exp} - \overline{x}_{\exp})^2} \qquad (5)$$

where $x_{\exp}$ are the experimental values, $\overline{x}_{\exp}$ are the means of the experimental values, and $x_{app}$ are the values obtained using the approximation formulas.

### 2.1 EA-model

This is the Edwards–Anderson model for a two-dimensional lattice where spins interact with their four nearest neighbors only and nonzero matrix elements are normally distributed.

We found that the function of Approximation functions derived as a result of analysis of our experimental data have the form

$$E_0 = 1.3151 - \frac{1.17}{L}, \quad \sigma_0 = \frac{0.74}{L} + \frac{0.11}{L^2}. \qquad (6)$$

The validities of these expressions are $R^2 = 0.994$ and $R^2 = 0.993$, respectively.

When comparing the expressions of Eq. (6) with the experiment, we see that these formulas describe them very well. In Fig. 1, we show that the function $E_0 = E_0(N)$ matches perfectly with data of Table 1. The value of the relative error $Err = 1 - E_0^{(\exp)} / E_0^{(approx)}$ is less than $2 \cdot 10^{-3}$.

The function $\sigma_0 = \sigma_0(N)$ (the second expression of Eq. (6)) in Fig. 2 also describes the data of Table 1 very well. The value of the relative error is less than 0.4%.

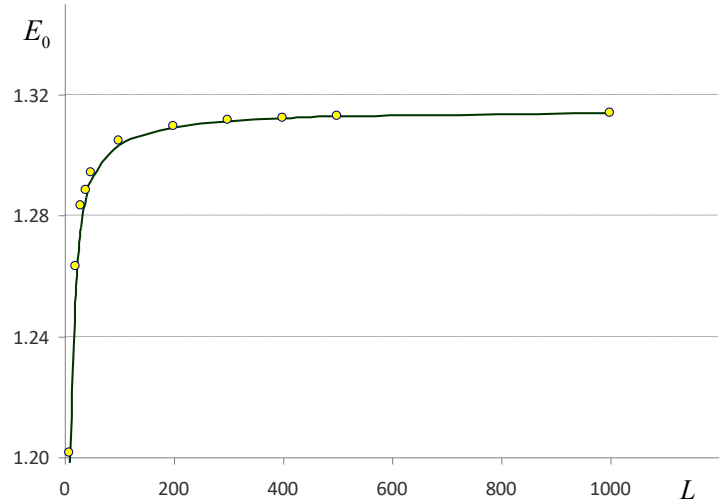

**Fig. 1.** $E_0$ vs $L$. EA-model: line – eqs.(6), circles – experiment.

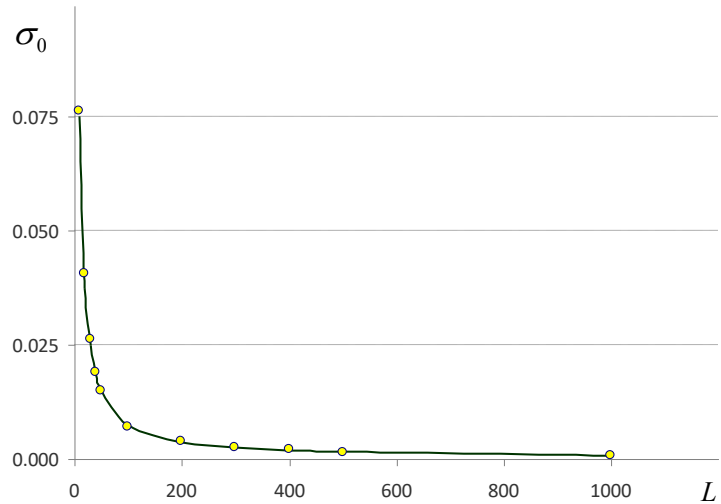

**Fig. 2.** $\sigma_0$ vs $L$. EA-model: line – eqs.(6), circles – experiment.

**Table 1.** Characteristics of global minima

|  |  | EA-model |  | EA*-model |  |
| --- | --- | --- | --- | --- | --- |
| $L$ | $M$ | $E_0$ | $\sigma_0$ | $E_0$ | $\sigma_0$ |
| 10 | 200 | 1.20245 | 0.07611 | 1.27347 | 0.05579 |
| 20 | 200 | 1.26305 | 0.04049 | 1.32793 | 0.02848 |
| 30 | 200 | 1.28416 | 0.02608 | 1.34307 | 0.01934 |
| 40 | 200 | 1.28838 | 0.01882 | 1.34864 | 0.01514 |
| 50 | 200 | 1.29569 | 0.01493 | 1.35738 | 0.01066 |
| 100 | 200 | 1.30448 | 0.00695 | 1.36681 | 0.00517 |
| 200 | 200 | 1.30957 | 0.00392 | 1.37173 | 0.00294 |
| 300 | 100 | 1.31136 | 0.00254 | 1.37337 | 0.00187 |
| 400 | 100 | 1.31229 | 0.00193 | 1.37467 | 0.00144 |
| 500 | 30 | 1.31279 | 0.00135 | 1.37498 | 0.00113 |
| 1000 | 30 | 1.31390 | 0.00074 | 1.37564 | 0.00054 |

## 2.2 EA*-model

This is the same Edwards–Anderson model for a two-dimensional lattice, but here the uniform distribution is used in place of the normal distribution.

In this case, approximation functions obtained after our analysis of the experimental data have the form

$$E_0 = 1.3769 - \frac{1.23}{L}, \quad \sigma_0 = \frac{0.54}{L} + \frac{0.2}{L^2} \qquad (7)$$

The validities of these expressions are $R^2 = 0.996$ and $R^2 = 0.992$, respectively.

Comparing the expressions of Eq. (7) with the experiment, we see that they describe it very well. In Fig. 3, we present the dependence $E_0 = E_0(N)$ (the first expression of Eq. (7)) that matches perfectly with the data from Table 1. When $L > 50$, the relative error is less than $2 \cdot 10^{-4}$.

The dependence $\sigma_0 = \sigma_0(N)$ (the second expression of Eq. (7)) shown in Fig. 4 also describes the data from Table 1 very well. Here the relative error is less than 0.5%.

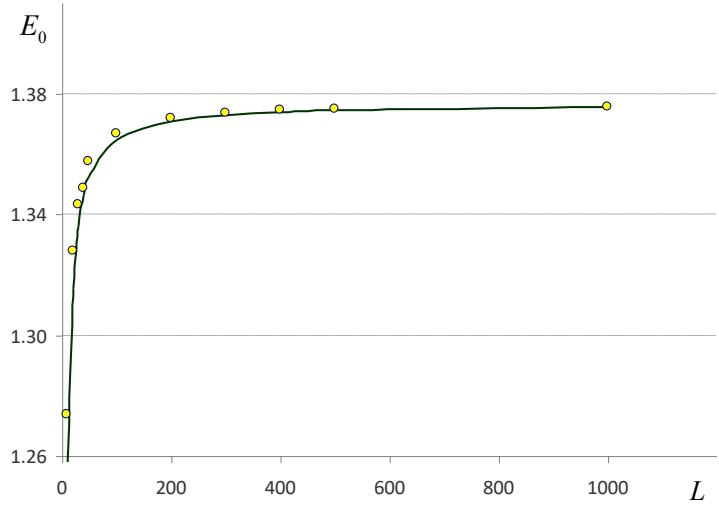

**Fig. 3.** $E_0$ vs $L$. EA-model: line – eqs.(6), circles – experiment.

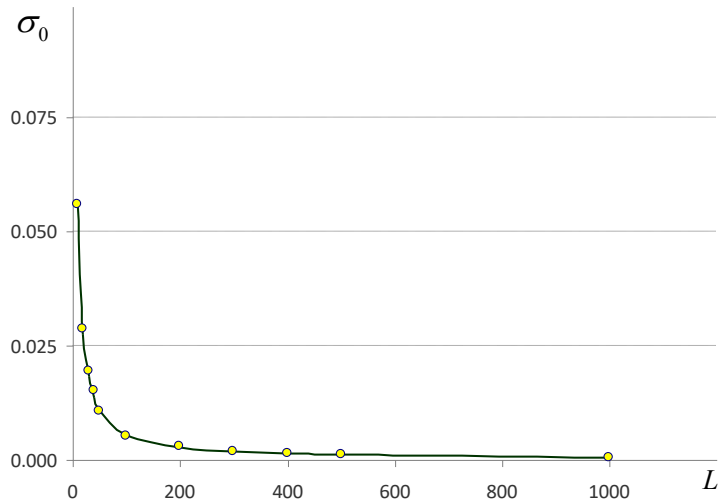

**Fig. 4.** $\sigma_0$ vs $L$. EA*- model: line – eqs.(7), circles – experiment.

## 3  Discussion

Our analysis of the two models allowed us to derive empirical relations in Eqs. (6) and (7) for the most important characteristics of the global minima (see Eqs. (6) and (7)). Our goal was to obtain expressions which with a high certainty described the dependences of these characteristics on $N$ in the whole range of the dimensions of the problem that we were able to examine. Based on these results, we had to determine the asymptotic behavior of these characteristics when $N \to \infty$. Evidently there are different approaches to approximation of the experimental data in Table 1. Consequently, it is possible to obtain a list of different expressions, and some of them can be even more accurate than the expressions of Eqs. (6) and (7). However, this fact does not change the goal of our study: independent of the form of the obtained approximation functions, they have to describe correctly the behavior of the characteristics inside the test range of $N$ and provide trustworthy asymptotic values when $N \to \infty$ (see Table 2).

**Table 2.** Asymptotic values of $E_0$ and $\sigma_0$ ($N \to \infty$).

|           | $E_0$             | $\sigma_0$ |
|-----------|-------------------|------------|
| EA-model  | $1.3151 \pm 0.002$ | $0.74/L$   |
| EA*-model | $1.3769 \pm 0.002$ | $0.54/L$   |

As we see, the data of Table 2 differ significantly from the values presented in Eq. (1). The point is that when minimizing the functional of Eq. (2) with a view to calculating $E_0$ different authors used different minimization algorithms. To do that, they defined $E_0$ as the energy corresponding to the deepest minimum, which under a reasonable number of tests frequently was far from $\bar{E}_0$. As an example, let us discuss the results of numerical experiments [40] in which they defined the energy of the deepest minimum $E^*$. Then, for the relative distance

$$\delta E = 100\% \cdot \frac{E_0 - E^*}{\bar{E}_0} \qquad (8)$$

between $E_0$ and $E^*$ when $L \geq 100$ we obtain:

$$\delta E = 16.45\% \pm 0.5\%, \qquad \text{for EA-model}, \qquad (9)$$

$$\delta E = 16.61\% \pm 0.4\%, \qquad \text{for EA*-model}. \qquad (10)$$

From our point of view, this is a possible reason why the estimates of $E_0$ obtained by different authors differ so significantly. Namely, when the size of the system is sufficiently large ($L \geq 30$) such an approach is not applicable since the probability of finding the global minimum in the course of a random search is exponentially small: it is $\sim \exp(-0.04N)$.

**Acknowledgements**

The work was supported by Russian Foundation for Basic Research (RFBR Project 18-07-00750).